\PassOptionsToPackage{bookmarks=true, colorlinks, citecolor=mycitecolor,linkcolor=mylinkcolor,urlcolor=mycitecolor}{hyperref}

\documentclass[conf]{new-aiaa}
\usepackage[utf8]{inputenc}

\usepackage{amsthm}
\usepackage{graphicx}
\usepackage{subcaption}
\usepackage{amsmath}
\usepackage{mathtools}
\usepackage[version=4]{mhchem}
\usepackage{siunitx}
\usepackage{longtable,tabularx}
\usepackage{algorithm}
\usepackage{algpseudocode}

\usepackage{xcolor}
\definecolor{mycitecolor}{RGB}{71, 191, 38}
\definecolor{mylinkcolor}{RGB}{40, 115, 201}

\newtheorem{theorem}{Theorem}

\newtheorem{definition}{Definition}

\newtheorem{problem}{Problem}

\title{Risk-Sensitive Orbital Debris Collision Avoidance using Distributionally Robust Chance Constraints}

\author{Kanghyun Ryu\footnote{PhD student, Mechanical Engineering Department, AIAA Student Member, kanghyun.ryu@berkeley.edu.}, Jean-Baptiste Bouvier\footnote{Postdoctoral Scholar, Mechanical Engineering Department, bouvier3@berkeley.edu.}, Shazaib Lalani\footnote{Undergraduate Student, Aerospace Engineering Program, AIAA Student Member, slalani29@berkeley.edu.}}
\affil{University of California Berkeley, Berkeley, CA, 94720}
\author{Siegfried Eggl\footnote{Assistant Professor, Aerospace Engineering Department, eggl@illinois.edu.}}
\affil{University of Illinois at Urbana-Champaign, Urbana, IL, 61801}
\author{Negar Mehr\footnote{Assistant Professor, Mechanical Engineering Department, negar@berkeley.edu.}}
\affil{University of California Berkeley, Berkeley, CA, 94720}

\begin{document}

\maketitle

\begin{abstract}
    The exponential increase in orbital debris and active satellites will lead to congested orbits, necessitating more frequent collision avoidance maneuvers by satellites. To minimize fuel consumption while ensuring the safety of satellites, enforcing a chance constraint, which poses an upper bound in collision probability with debris, can serve as an intuitive safety measure. However, accurately evaluating collision probability, which is critical for the effective implementation of chance constraints, remains a non-trivial task. This difficulty arises because uncertainty propagation in nonlinear orbit dynamics typically provides only limited information, such as finite samples or moment estimates about the underlying arbitrary non-Gaussian distributions. Furthermore, even if the full distribution were known, it remains unclear how to effectively compute chance constraints with such non-Gaussian distributions.
    To address these challenges, we propose a distributionally robust chance-constrained collision avoidance algorithm that provides a sufficient condition for collision probabilities under limited information about the underlying non-Gaussian distribution. Our distributionally robust approach satisfies the chance constraint for all debris position distributions sharing a given mean and covariance, thereby enabling the enforcement of chance constraints with limited distributional information. To achieve computational tractability, the chance constraint is approximated using a Conditional Value-at-Risk (CVaR) constraint, which gives a conservative and tractable approximation of the distributionally robust chance constraint. We validate our algorithm on a real-world inspired satellite-debris conjunction scenario with different uncertainty propagation methods and show that our controller can effectively avoid collisions.
\end{abstract}

\section{Nomenclature}

{\renewcommand\arraystretch{1.0}
\noindent\begin{longtable*}{@{}l @{\quad=\quad} l@{}}
$\mathbf{a}_{drag}$ & atmospheric drag \\
$A$ & effective area of the object for atmospheric drag \\
$C_d$ & drag coefficient of the object \\
$m$ & mass of the object \\
$r_0$ & sea level altitude for the atmospheric model \\
$\mu_E$ & standard gravitational parameter of the Earth \\
$\rho_0$ & atmospheric pressure at sea level \\
$\boldsymbol{\omega}_E$ & rotational velocity of the Earth \\
$\mathbf{r}$ & position of the object in ECI frame\\
$\mathbf{v} = \dot{\mathbf{r}}$ & velocity of the object in ECI frame\\
$\mathbf{x} = [\mathbf{r}, \mathbf{v}]$ & full state of the object in ECI frame\\
$\mathcal{R}_{free}$ & a set of debris positions where debris is collision-free with the satellite \\
$\varepsilon$ & allowable probability of collision \\
$VaR$ & Value-at-Risk \\
$CVaR$ & Conditional Value-at-Risk \\
$Tr\{\cdot\}$ & matrix trace operator \\
$chol(\cdot)$ & Cholesky decomposition operator
\end{longtable*}}

\section{Introduction}

Over the past two decades, the exponential increase in operational satellites, coupled with a significant accumulation of space debris, residuals from defunct satellites or previous space missions, has started to increase collision risk, particularly in Low Earth Orbits~\cite{space_debris}. This intensifying orbital congestion poses substantial risks not only to the functional integrity of satellites~\cite{CNN_ISS_debris}, but also to the safety and sustainability of future space initiatives~\cite{debris_risk}. Consequently, the development of an advanced satellite collision avoidance algorithm, capable of intelligently predicting potential collisions and autonomously adjusting satellite orbits, is imperative~\cite{braun2016operational}. 

In collision avoidance maneuvers, accounting for uncertainties in orbit propagation is paramount since accurately forecasting the orbital position and velocity of approaching space debris comes with considerable challenges~\cite{luo2017review}. These uncertainties arise primarily due to inaccuracies inherent in state estimation and dynamics models. Therefore, collision avoidance maneuvers must be - and generally are - based on accurate risk assessment. The latter generally accounts for uncertainties associated with space debris trajectories as well as the potential severity of a collision event. 

In this work, we formulate the collision avoidance problem under debris position uncertainty as a chance-constrained optimal control problem, enforcing an upper bound on the collision probability. Accurate estimation of obstacle uncertainty is crucial for computing the chance constraint, which required accurately propagating uncertainty under complex orbital dynamics. Previous studies on chance-constrained collision avoidance maneuvers have relied on the assumption that orbit covariances remain Gaussian~\cite{dutta2022convex, chan2008spacecraft}. However, in highly nonlinear orbital dynamics, even if the initial uncertainty is Gaussian, it tends to evolve rapidly into a non-Gaussian distribution due to the system's inherent nonlinearity~\cite{giza2009approach}. While other methods as the Unscented Transform~\cite{vishwajeet2014nonlinear, adurthi2015conjugate} have been also proposed for orbital uncertainty propagation, they typically provide only limited information about the uncertainty distribution, such as the first two moments, rather than providing the full probability density function required for precise evaluation of chance constraints.

Even for known distributions, evaluating chance constraints for non-Gaussian uncertainty becomes computationally intractable~\cite{nemirovski2007convex}. Therefore, evaluating chance constraints with non-Gaussian uncertainty typically relies on Monte Carlo simulations to evaluate the ratio of samples encountering a collision compared to the total number of samples~\cite{de2010monte, sabol2011probability}. However, this approach requires a large sample size for accurate approximation of the posterior distributions. Specifically, the required sample size for chance constraint approximation is proportional to the log of the inverse of the collision probability~\cite{dagum2000optimal}, which makes this method impractical for evaluating collision in orbit, where collision avoidance guidelines often require collision probability to be very small. 

To overcome these challenges, we propose a distributionally robust chance-constrained collision avoidance algorithm that aims to satisfy upper bounds in collision probability under limited information about the uncertainty distribution in a computationally efficient way. Our controller extends chance constraints to distributionally robust chance constraints, ensuring the satisfaction of the chance constraint for all debris uncertainty distributions within a specified ambiguity set~\cite{rahimian2019distributionally}. This ambiguity set covers all distributions sharing the same mean and covariance, including non-Gaussian distributions which can arise in nonlinear orbit dynamics. Ensuring distributionally robust chance constraints over this ambiguity set results in a conservative approximation of the chance constraints when only mean and covariance information is available. Consequently, our controller is agnostic to the uncertainty propagation method, provided that the mean and covariance information can be obtained, which offers flexibility in choosing the uncertainty propagation approach. To make evaluating the distributionally robust chance constraint computationally tractable, we leverage concepts from risk-sensitive control~\cite{chow2015risk}. Specifically, we demonstrate the equivalence between the chance constraint and Value-at-Risk (VaR), which ultimately leads to a conservative, computationally tractable Conditional Value-at-Risk (CVaR) approximation~\cite{van2015distributionally, ryu2024integrating}.

To validate that our controller is agnostic to the choice of uncertainty propagation method, we implement it with three different approaches: a Linear Gaussian uncertainty propagator~\cite{dutta2022convex}, an Unscented Transform uncertainty propagator~\cite{vishwajeet2014nonlinear}, and a Monte Carlo uncertainty propagator~\cite{de2010monte}. We evaluate our controller in a real-world-inspired~\cite{kelso2006socrates} close approach scenario and demonstrate that it consistently maintains a minimum safe distance from the debris, thereby avoiding collisions effectively.

To summarize, our main contributions in this work are as follows.
\begin{enumerate}
    \item We design a distributionally robust chance-constrained collision avoidance controller that is agnostic to the uncertainty propagation method requiring only the mean and covariance of the debris's position distribution.
    \item Our controller provides a conservative approximation of the collision avoidance chance constraint that is computationally tractable for non-Gaussian uncertainty.
    \item We validate our controller with different uncertainty propagation methods and show that our controller can successfully avoid collision. 
\end{enumerate}

The remainder of this paper is structured as follows. Section~\ref{sec: related works} reviews related works on uncertainty propagation and collision avoidance under uncertainty. We precisely formulate our problem of interest in Section~\ref{sec: problem}. Section~\ref{sec: preliminaries} introduces the concept of distributionally robust constraints and the approximation of chance constraints by CVaR constraints. The details of our proposed approach are described in Section~\ref{sec: method}. Section~\ref{sec: experiment settings} outlines our experimental setup, and the simulation results are presented in Section~\ref{sec: results}. Finally, Section~\ref{sec: conclusion} summarizes this paper and discusses future directions of work.

\section{Related Works}\label{sec: related works}

\subsection{Orbit Uncertainty Propagation}

As consideration of uncertainty is necessary for safe and efficient maneuvers in complex in-orbit scenarios, numerous works have investigated uncertainty propagation in the orbit~\cite{luo2017review}. While Monte Carlo simulations often provide accurate uncertainty propagation~\cite{junkins1996non}, they require a large number of samples to provide statistical guarantees, which makes them computationally inefficient. A common alternative to model uncertainties in orbital mechanics is to use Gaussian probability distributions with linear uncertainty propagation~\cite{dutta2022convex}. This approach is suitable for onboard autonomous navigation thanks to its simplicity and computational efficiency. However, such a linear uncertainty propagation accumulates significant error over long-horizon propagation of the highly nonlinear spaceflight dynamics~\cite{vavilov2020partial}. Especially, long-term uncertainty tends to be curved and stretched along the object's nominal orbit~\cite{vavilov2020partial}. To address these challenges, several works have proposed nonlinear uncertainty propagation methods including polynomial chaos expansions~\cite{jones2013nonlinear}, state transition tensors~\cite{lee2016analytical}, and Gaussian mixture models~\cite{vittaldev2016space}. Still, integrating these nonlinear uncertainty propagators with collision avoidance maneuvers is a complex task since they tend to give limited information about the distribution or pose certain assumptions about the distribution. In this work, we utilize distributionally robust chance constraints that can provide collision probability bound while being agnostic to the uncertainty propagation method.

\subsection{Collision Avoidance Maneuvers under Uncertainty}
Uncertainties in the dynamics and states are important factors in the planning of collision avoidance maneuvers involving space debris. Within the earlier works on collision avoidance under uncertainty,~\cite{slater2006} employs a stochastic model to calculate collision likelihood, propagating uncertainties in initial conditions, dynamic perturbations, and state estimation through linearized equations of motion. This method minimizes collision risks while conserving fuel through energy-efficient maneuvers informed by the evolving uncertainties. Building on this approach, our work integrates a distributionally robust constraint, maintaining collision probability bounds and optimizing fuel use while addressing potentially non-Gaussian uncertainty distributions beyond linearized models. Similarly,~\cite{dolan2023satellite} introduces a game-theory-based strategy for satellite collision avoidance, treating interactions as a repeated game. Satellites adopt probabilistic strategies to maneuver or hold course, dynamically adjusting decisions based on collision probabilities and time to closest approach. While this method excels in operator-operator settings, our approach focuses on collision with space debris, where only one active player is capable of modifying the interaction between objects. Another work~\cite{9483158} develops a stochastic optimal control framework for multi-satellite collision avoidance, utilizing non-convex, probabilistic constraints and risk allocation via difference-of-convex programming. Instead, our approach enforces collision probability by employing a distributionally robust method that is agnostic to specific uncertainty propagation models, enabling scalability for unknown uncertainty distributions. Finally, ~\cite{dutta2022convex} proposes a convex optimization approach using Gaussian distributions to model collision risks, integrating Mahalanobis distance and instantaneous collision probability. In contrast, we adopt a distributionally robust approach and a CVaR approximation, accommodating a broader range of uncertainty distributions over Gaussian assumptions.

\section{Problem Formulation} \label{sec: problem}

In this section, we formally define our collision avoidance problem. We consider an orbital debris avoidance scenario in low Earth orbit. While satellite dynamics and state estimation are assumed to be deterministic which is motivated by the fact that most active satellites have GNSS trackers on board, we will take into account that there is uncertainty in debris dynamics and initial state estimation, and we aim to ensure an upper bound on the collision probability under these uncertainties. 

\subsection{Satellite and Obstacle Dynamics}

We use Cartesian coordinates in the Earth-Centered Inertial frame to model the general nonlinear dynamics of the objects, including the considered satellite and space debris. We use subscript $s$ and $d$ to denote variables associated with the satellite and debris respectively. We denote the space object's position vector as $\mathbf{r}$, velocity vector as $\mathbf{v}$, and state vector as $\mathbf{x} = [\mathbf{r}^T, \mathbf{v}^T]^T$. We denote vector variables using bold symbols, such as the satellite's position vector $\mathbf{r}_s$, while we denote their Euclidean norms using non-bold symbols, e.g., $r_s = ||\mathbf{r}_s||$. Following~\cite{dutta2022convex}, the satellite dynamics are as follows
\begin{equation}\label{eqn: satellite dynamics}
    \dot{\mathbf{x}}_s = 
    \begin{bmatrix}
        \dot{\mathbf{r}}_s \\
        \dot{\mathbf{v}}_s
    \end{bmatrix}
    = 
    \begin{bmatrix}
        \mathbf{v}_s \\
        -\frac{\mu_E}{r_s^3} \mathbf{r}_s + \mathbf{a}_{s, drag}
    \end{bmatrix}
    + 
    \begin{bmatrix}
        0 \\
        \mathbf{u}_s
    \end{bmatrix}
\end{equation}
where $\mu_E$ is standard gravitational parameter of the Earth, $\mathbf{a}_{drag}$ is atmospheric drag, and $\mathbf{u}_s$ is control input for the satellite.

The debris is not actuated and hence we assume that it follows the same dynamics as the spacecraft with the exception of the control term $\mathbf{u}$. Furthermore, since the exact dynamics of the debris may not be perfectly known, we include system uncertainties, denoted by $\mathbf{w}_d$.
\begin{equation}\label{eqn: debris dynamics}
    \dot{\mathbf{x}}_d = 
    \begin{bmatrix}
        \dot{\mathbf{r}}_d \\
        \dot{\mathbf{v}}_d
    \end{bmatrix}
    = 
    \begin{bmatrix}
        \mathbf{v}_d \\
        -\frac{\mu_E}{r_d^3} \mathbf{r}_d + \mathbf{a}_{d, drag}
    \end{bmatrix}
    + \mathbf{w}_d.
\end{equation}



In both satellite~\eqref{eqn: satellite dynamics} and debris dynamics~\eqref{eqn: debris dynamics}, the only perturbation considered is atmospheric drag. However, other perturbative forces, such as J2 perturbation~\cite{vadali2009model}, can also be incorporated. We model atmospheric drag as
\begin{equation*}
    \mathbf{a}_{drag} = - \frac{1}{2} \rho(r) \frac{A C_d}{m}v_{0} \mathbf{v}_{0}
\end{equation*}
with $\mathbf{v}_0 = \mathbf{v} - \boldsymbol{\omega}_E \times \mathbf{r}$, where $\rho(r)$ is atmosphere density at altitude $r$, $A$ is effective area of the object for atmospheric drag, $C_d$ is drag coefficient, $m$ is the mass of the object, and $\boldsymbol{\omega}_E$ is rotational velocity of the Earth. We employ the exponential atmosphere model to predict atmosphere density at altitude $r$
\begin{equation*}
    \rho(r) = \rho_0\ exp\Big(-\frac{r-r_0}{r_0}\Big)
\end{equation*}
where $r_0$ is the sea level altitude for the atmospheric model.

As we need discrete-time dynamics for the numerical simulation, we use the fourth-order Runge-Kutta (RK4) method for discretizing the continuous-time dynamics. Then, we can denote the discrete-time dynamics model of the satellite and the debris as
\begin{subequations}\label{eqn: discrete dynamics}
    \begin{align}
        \mathbf{x}_{s}^{k+1} &= f_s(\mathbf{x}_{s}^{k}, \mathbf{u}_{s}^{k}) \label{eqn: discrete satellite dynamics}\\ 
        \mathbf{x}_{d}^{k+1} &= f_d(\mathbf{x}_{d}^{k}, \mathbf{w}_{d}^{k}) \label{eqn: discrete debris dynamics}
    \end{align}
\end{subequations}
where superscript $k$ denotes the time step.

\subsection{Model Predictive Control with Collision Avoidance Chance Constraints}\label{subsec: mpc formulation}

\begin{figure}
    \centering
    \includegraphics[width=0.5\linewidth]{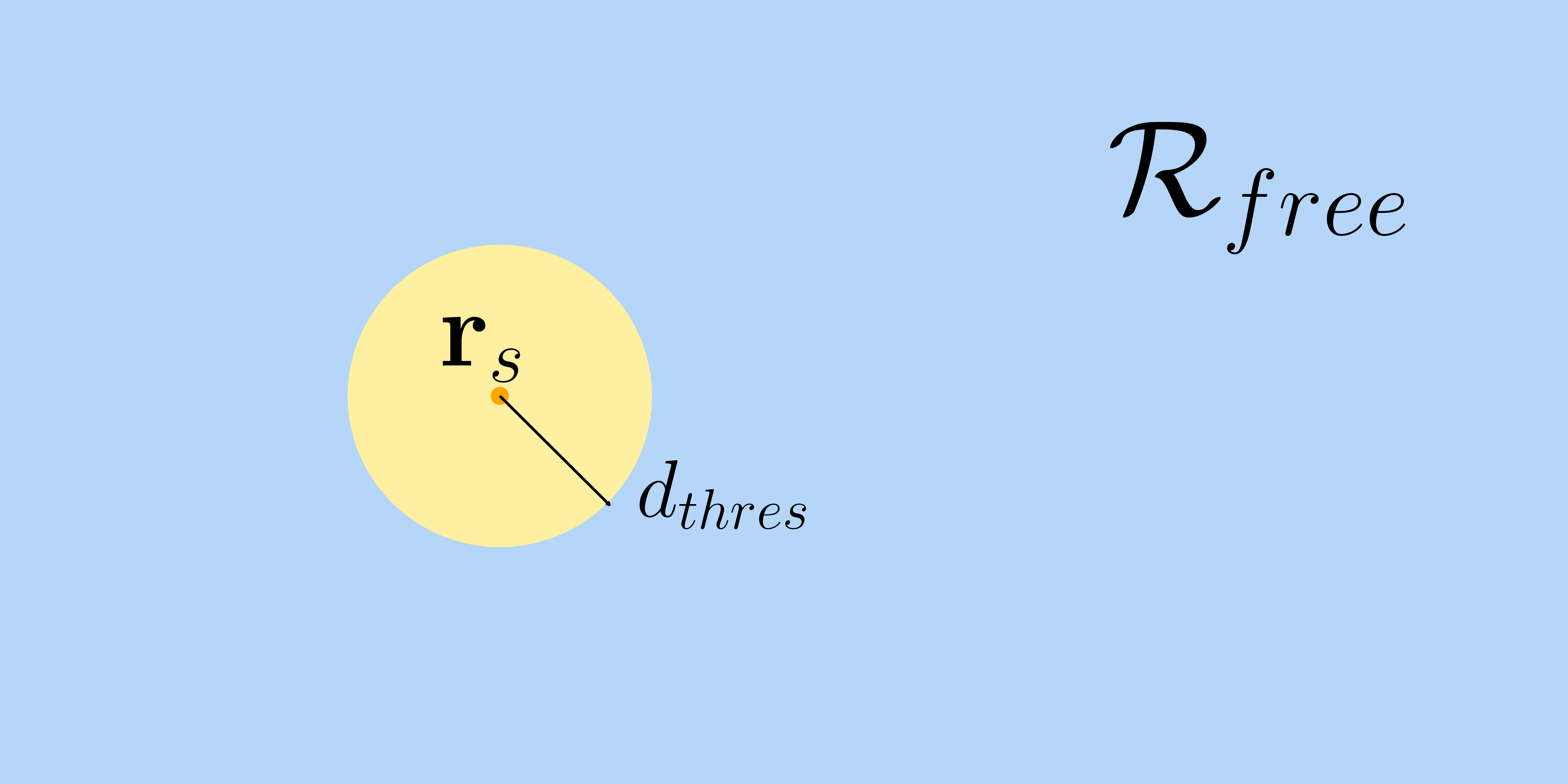}
    \caption{Based on the satellite position $\mathbf{r}_s$, the region outside of the collision threshold $d_{thres}$ is defined as the collision-free set $\mathcal{R}_{free}$ for debris.}
    \label{fig:original safe set}
\end{figure}

While conventional worst-case robust constraints~\cite{mueller2008collision, mueller2009onboard} can provide hard safety guarantees, they are not suitable for uncertainties with unbounded support such as orbital uncertainty. Therefore, we use chance constraints to enforce an upper bound on the collision probability under unbounded support of the uncertainty. We consider that a collision occurs when the distance between the satellite and the debris becomes smaller than a certain threshold $d_{thres}$. Then, we can define the system as safe if the debris' position is within the collision-free set $\mathcal{R}_{free}$, as illustrated in Figure~\ref{fig:original safe set}.
\begin{equation}\label{eqn: general collision free set}
    \mathcal{R}_{free} := \{\mathbf{r}_d \in \mathbb{R}^3 : ||\mathbf{r}_d - \mathbf{r}_s || > d_{thres} \}
\end{equation}

With this definition, we can define our collision avoidance chance constraint as follows.
\begin{definition}[Chance constraints for collision avoidance]
    For a random position vector for debris $\mathbf{r}_d \in \mathbb{R}^3$ whose underlying distribution is $\mathbb{P}^*$, we require 
    \begin{equation}\label{eqn: chance constraint def}
        \textnormal{Prob}^{\mathbb{P}^*}(\mathbf{r}_d \in \mathcal{R}_{free}) \geq 1 - \varepsilon,
    \end{equation}
    where $\varepsilon$ is the upper bound of the collision probability.
\end{definition}

This chance constraint enforces collision probability under $\varepsilon$, providing intuitive safety measures for satellite navigation. Then, we formulate our collision avoidance maneuver as a fuel-optimal model predictive control (MPC) problem with chance constraints to enforce the upper bound in collision probability. MPC objective is to minimize fuel consumption cost $\mathcal{J}$ over a receding-horizon $K$ assuming that the initial debris state follows Gaussian distribution with mean $\Bar{\mathbf{x}}_d^0$ and covariance $P_d^0$. While we model initial uncertainty for the debris state as Gaussian, the future debris state uncertainty quickly diverges from Gaussian due to its nonlinear dynamics, which makes collision avoidance chance constraint computationally intractable.

\begin{problem} \label{problem:constrained mpc}
    Given the satellite dynamics $f_s$~\eqref{eqn: discrete satellite dynamics}, debris dynamics $f_d$~\eqref{eqn: discrete debris dynamics}, and uncertainty in debris' initial state $\mathcal{N}(\Bar{\mathbf{x}}_d^0, P_d^0)$, we solve a receding horizon optimal control problem with a collision avoidance chance constraint~\eqref{eqn: chance constraint def}
    \begin{subequations}
        \begin{align}
            \min_{\mathbf{u}_s^{0:K-1}} \mathcal{J} &:= \sum_{k\, =\, 0}^{K-1}  ({\mathbf{u}_s^k})^T R (\mathbf{u}_s^k) \\
            s.t. \quad & \mathbf{x}_{s}^{k+1} = f_s(\mathbf{x}_{s}^{k}, \mathbf{u}_{s}^{k}), \\
                & \mathbf{x}_{d}^{k+1} = f_d(\mathbf{x}_{d}^{k}, \mathbf{w}_{d}^{k}), \\
                & \mathbf{x}_{d}^{0} \sim \mathcal{N}(\Bar{\mathbf{x}}_d^0, P_d^0) \label{eqn: problem initial uncertainty} \\
               & \textnormal{Prob}^{\mathbb{P}^*} (\mathbf{r}_d^k \in \mathcal{R}_{free}^k) \geq 1-\varepsilon \quad \forall~ 1 \leq k \leq K, \label{eqn: problem chance constraint}
        \end{align}
    \end{subequations}
    where $\mathbf{u}_s^{0:K-1} := (\mathbf{u}_s^0, \dots, \mathbf{u}_s^{K-1})$ are the satellite thrust inputs over horizon $K$, and $R$ is a positive definite matrix that assigns different weights to the components of $\mathbf{u}_s$.
\end{problem}

\section{Preliminaries} \label{sec: preliminaries}

In this section, we give mathematical preliminaries for how to evaluate chance constraints when there is limited information about the underlying non-Gaussian uncertainty associated with debris position. First, we show that distributionally robust chance constraints can be a sufficient condition for the satisfaction of a chance constraint when only limited information for the debris state distribution is given. Then, we introduce relationships between chance constraints with Value-at-Risk (VaR) and Conditional Value-at-Risk (CVaR), which gives tractable solutions of distributionally robust chance constraints.

\subsection{Distributionally Robust Chance Constraints}

While chance constraints provide an intuitive measure of safety, evaluating chance constraint \eqref{eqn: chance constraint def} requires an exact characterization of the probability distribution of the debris location $\mathbb{P}^*$. This requirement is challenging for in-orbit collision avoidance scenarios for several reasons. First, nonlinear orbit dynamics make analytical uncertainty propagation using a linearized model diverge quickly from the true uncertainty~\cite{giza2009approach}. Moreover, while Monte Carlo simulation can be applied to consider nonlinear orbit dynamics and non-Gaussian uncertainty, it requires too many samples to accurately capture the probability density function of the uncertainty. Finally, while several other methods such as unscented transform~\cite{vishwajeet2014nonlinear} or curvilinear coordinate system~\cite{vavilov2020partial} can be applied for nonlinear uncertainty propagation, they provide limited information about moments of distributions rather than the exact model. 

To address this issue, we adopt a distributionally robust approach where we aim to satisfy chance constraint~\eqref{eqn: chance constraint def} not only for a single distribution but for a group of distributions termed the \emph{ambiguity set} $\mathcal{P}$. Under the assumption that true distribution $\mathbb{P}^*$ is in the ambiguity set $\mathcal{P}$, distributionally robust chance constraint yields a conservative approximation of the chance constraint.
\begin{align}\label{eqn: distributionally robust relationships}
    \inf_{\mathbb{P} \, \in \, \mathcal{P}} \textnormal{Prob}^{\mathbb{P}} (\mathbf{r}_d \in \mathcal{R}_{free}) \geq 1-\varepsilon \Rightarrow \textnormal{Prob}^{\mathbb{P}^*}(\mathbf{r}_d \in \mathcal{R}_{free}) \geq 1-\varepsilon \quad \textnormal{if } ~ \mathbb{P}^* \in \mathcal{P}.
\end{align}

Therefore, when only limited information about the distribution is available, we can enforce the chance constraint using a distributionally robust chance constraint by constructing an ambiguity set that encompasses the true distribution based on the available information.

\subsection{Chance Constraints and their CVaR Approximations}

While a distributionally robust chance constraint gives a conservative and intuitive measure of safety when only limited information about the distribution is available, evaluating distributionally robust chance constraints \eqref{eqn: distributionally robust relationships} remains computationally intractable. In fact, evaluating chance constraint \eqref{eqn: chance constraint def} for a single distribution remains computationally intractable for non-Gaussian distributions~\cite{nemirovski2007convex}, as it requires the computation of multi-dimensional integral of an arbitrary probability density function. Moreover, a feasible set of chance-constrained problems is typically non-convex and disconnected, making it difficult to find a solution to the chance-constrained optimization problem~\cite{zymler2013distributionally}. Therefore, we approximate the chance constraint as a CVaR constraint which is known as a tight convex approximation of chance constraint~\cite{hong2014conditional}. Before defining the CVaR, we introduce the safety cost which is a continuous and smooth function that can represent a collision-free set $\mathcal{R}_{free}$.

\begin{definition}[Safety cost]
    We assume that we can characterize collision-free set $\mathcal{R}_{free}$~\eqref{eqn: general collision free set} as a sublevel set of a safety cost function $l(\mathbf{r}_d)$:
    \begin{equation}\label{eqn: loss function def}
        \mathcal{R}_{free} = \big\{\mathbf{r}_d \in \mathbb{R}^3\ |\ l(\mathbf{r}_d) \leq 0 \big\} \quad \textnormal{where } l: \mathbb{R}^3 \rightarrow \mathbb{R}.
    \end{equation}
\end{definition}

Therefore, positive safety cost indicates that there is potential for a collision between the satellite and debris, while negative safety cost means their configuration is collision-free. Then, we can re-write our chance constraint for collision avoidance \eqref{eqn: chance constraint def} using the safety cost function as
\begin{equation}\label{eqn: chance constraint}
    \textnormal{Prob}^{\mathbb{P}^*}(l(\mathbf{r}_d) \leq 0) \geq 1 - \varepsilon.
\end{equation}
This safety cost provides a continuous mapping from the position of debris to a scalar value that quantifies the potential risk for collision. Based on this safety cost, we can translate our chance constraint~\eqref{eqn: chance constraint} to the Value-at-Risk (VaR) constraint.

\begin{definition}[Value-at-Risk (VaR)]
    For a random position vector of debris $\mathbf{r}_d \in \mathbb{R}^3$ with true probability distribution $\mathbb{P}^*$ and safety cost function $l(\cdot)$ defined as \eqref{eqn: loss function def}, the VaR related to the safety cost distribution $l(\mathbf{r}_d)$ is defined as:
    \begin{equation*}
        \textnormal{VaR}_{\varepsilon}^{\mathbb{P}^*}\big( l(\mathbf{r}_d) \big) \coloneqq \inf \big\{ \gamma \in \mathbb{R} | \textnormal{Prob}^{\mathbb{P}^*}(l(\mathbf{r}_d) > \gamma) \leq \varepsilon \big\}.
    \end{equation*}    
\end{definition}

The definition of VaR shows that the probability of having a safety cost larger than VaR should be less than $\varepsilon$, i.e., $\textnormal{Prob}^{\mathbb{P}^*}(l(\mathbf{r}_d) > \textnormal{VaR}_{\varepsilon}^{\mathbb{P}^*}) \leq \varepsilon$. Therefore, following~\cite{van2015distributionally}, we can easily find an equivalence between chance constraint \eqref{eqn: chance constraint} and VaR as follows
\begin{equation*}
    \textnormal{VaR}_{\varepsilon}^{\mathbb{P}^*}\big( l(\mathbf{r}_d) \big) \leq 0 \Leftrightarrow \textnormal{Prob}^{\mathbb{P}^*}(l(\mathbf{r}_d) > 0) \leq \varepsilon \Leftrightarrow \textnormal{Prob}^{\mathbb{P}^*}(l(\mathbf{r}_d) \leq 0) \geq 1 - \varepsilon.
\end{equation*}
Evaluating the distributionally robust version of the VaR constraint is still intractable as VaR remains a non-convex risk measure for $l(\mathbf{r}_d)$~\cite{van2015distributionally}. To address this, we approximate the VaR constraint with a Conditional Value-at-Risk (CVaR) constraint which is known as a convex conservative approximation of VaR.

\begin{definition}[Conditional Value-at-Risk (CVaR)]
    For a random position vector of debris $\mathbf{r}_d \in \mathbb{R}^3$ with true probability distribution $\mathbb{P}^*$ and safety cost function $l(\cdot)$~\eqref{eqn: loss function def}, the CVaR of safety cost related to that position distribution is defined as:
    \begin{equation} \label{eqn: CVaR definition}
        \textnormal{CVaR}_{\varepsilon}^{\mathbb{P}^*}\big( l(\mathbf{r}_d) \big) \coloneqq \inf_{\beta \in \mathbb{R}} \Bigg\{\beta + \frac{1}{\varepsilon} \mathbb{E}_{\mathbb{P}^*}\big\{ (l(\mathbf{r}_d)-\beta)^+ \big\} \Bigg\}, \quad \text{where} \quad (\cdot)^+ = \max \{\cdot, 0\}.
    \end{equation}
\end{definition}

It is known that \eqref{eqn: CVaR definition} is equivalent to a conditional expectation of safety cost above VaR \cite{rockafellar2000optimization}.
\begin{equation*}
    \textnormal{CVaR}_{\varepsilon}^{\mathbb{P}^*}\big( l(\mathbf{r}_d) \big) = \mathbb{E}_{\mathbb{P}^*} \big[ l(\mathbf{r}_d) > \textnormal{VaR}_{\varepsilon}^{\mathbb{P}^*}\big( l(\mathbf{r}_d) \big].
\end{equation*}
Because of this conditional expectation property, CVaR is always larger than or equal to VaR. Thus, we can show that the CVaR constraint is a sufficient condition for satisfying the chance constraint:
\begin{align} \label{eqn:approximation}
    \textnormal{CVaR}_{\varepsilon}^{\mathbb{P}^*}\big( l(\mathbf{r}_d) \big) \leq 0 \Rightarrow \textnormal{VaR}_{\varepsilon}^{\mathbb{P}^*}\big( l(\mathbf{r}_d) \big) \leq 0 \Leftrightarrow 
    \textnormal{Prob}^{\mathbb{P}^*}(\mathbf{r}_d \in \mathcal{R}_{free}) \geq 1-\varepsilon.
\end{align}
Finally, this relationship between CVaR and chance constraint can be extended to their distributionally robust counterparts:
\begin{align} \label{eqn: distributionally robust cvar approximation}
    \sup_{\mathbb{P} \, \in \, \mathcal{P}} \textnormal{CVaR}_\varepsilon^{\mathbb{P}} (l(\mathbf{r}_d)) \leq 0 
    \Rightarrow \sup_{\mathbb{P} \, \in \, \mathcal{P}} \textnormal{VaR}_{\varepsilon}^{\mathbb{P}^*}\big( l(\mathbf{r}_d) \big) \leq 0
    \Leftrightarrow \inf_{\mathbb{P} \, \in \, \mathcal{P}} \textnormal{Prob}^{\mathbb{P}} (l(\mathbf{r}_d) \leq 0) \geq 1-\varepsilon.
\end{align}

\section{Risk-Sensitive Collision Avoidance using Distributionally Robust Chance Constraints} \label{sec: method}

In this section, we will introduce how we utilize distributionally robust chance constraints to guarantee the satisfaction of the chance constraint when only the mean and covariance of debris position distribution are given. We show that enforcing the distributionally robust CVaR constraint gives closed-form sufficient conditions for enforcing the distributionally chance constraint. Finally, we provide details on how we solve the reformulated optimal control problem using sampling-based optimization methods.

\subsection{Reformulation of Chance Constraint by Distributionally Robust CVaR Constraints}
Using the relationship between distributionally robust CVaR constraint and chance constraint~\eqref{eqn: distributionally robust cvar approximation}, we aim to approximate chance constraint~\eqref{eqn: problem chance constraint} in Problem~\ref{problem:constrained mpc} by its relevant distributionally robust CVaR constraint.
More specifically, we construct a moment-based ambiguity set for debris position $\mathbf{r}_d^k$ at time $k$ as the set of all distributions sharing the same estimated mean $\mu_d^k$ and covariance $\Sigma_d^k$ obtained from an uncertainty propagation method.

\begin{equation}\label{eqn: moment based ambiguity set}
    \mathcal{P}^k = \big\{ \mathbb{P} ~ | ~ \mathbb{E}_{\mathbb{P}} [\mathbf{r}_d^k] = \mu_d^k, ~~\mathbb{E}_{\mathbb{P}} [(\mathbf{r}_d^k - \mu_d^k)(\mathbf{r}_d^k - \mu_d^k)^T] = \Sigma_d^k \big\}.
\end{equation}
We chose a moment-based ambiguity set since many analytical uncertainty propagation methods, such as linear Gaussian propagation~\cite{dutta2022convex} or unscented transform~\cite{vishwajeet2014nonlinear}, can provide estimations of the mean and covariance. With this assumption in mind, distributionally robust CVaR constraint provides sufficient conditions for satisfying the chance constraint under limited knowledge about the distribution, relying solely on the mean and covariance of the position distribution:
\begin{equation*}
    \sup_{\mathbb{P} \, \in \, \mathcal{P}} \textnormal{CVaR}_\varepsilon^{\mathbb{P}}  (l^k(\mathbf{r}_d^k)) \leq 0 \Rightarrow \textnormal{Prob}^{\mathbb{P}^*}(\mathbf{r}_d^k \in \mathcal{R}_{free}^k) \geq 1-\varepsilon \quad \textnormal{if }~ \mathbb{E}_{\mathbb{P}^*} [\mathbf{r}_d^k] = \mu_d^k, ~~\mathbb{E}_{\mathbb{P}^*} [(\mathbf{r}_d^k - \mu_d^k)(\mathbf{r}_d^k - \mu_d^k)^T] = \Sigma_d^k.
\end{equation*}
\begin{figure}[t]
    \centering
    \includegraphics[width=0.4\linewidth]{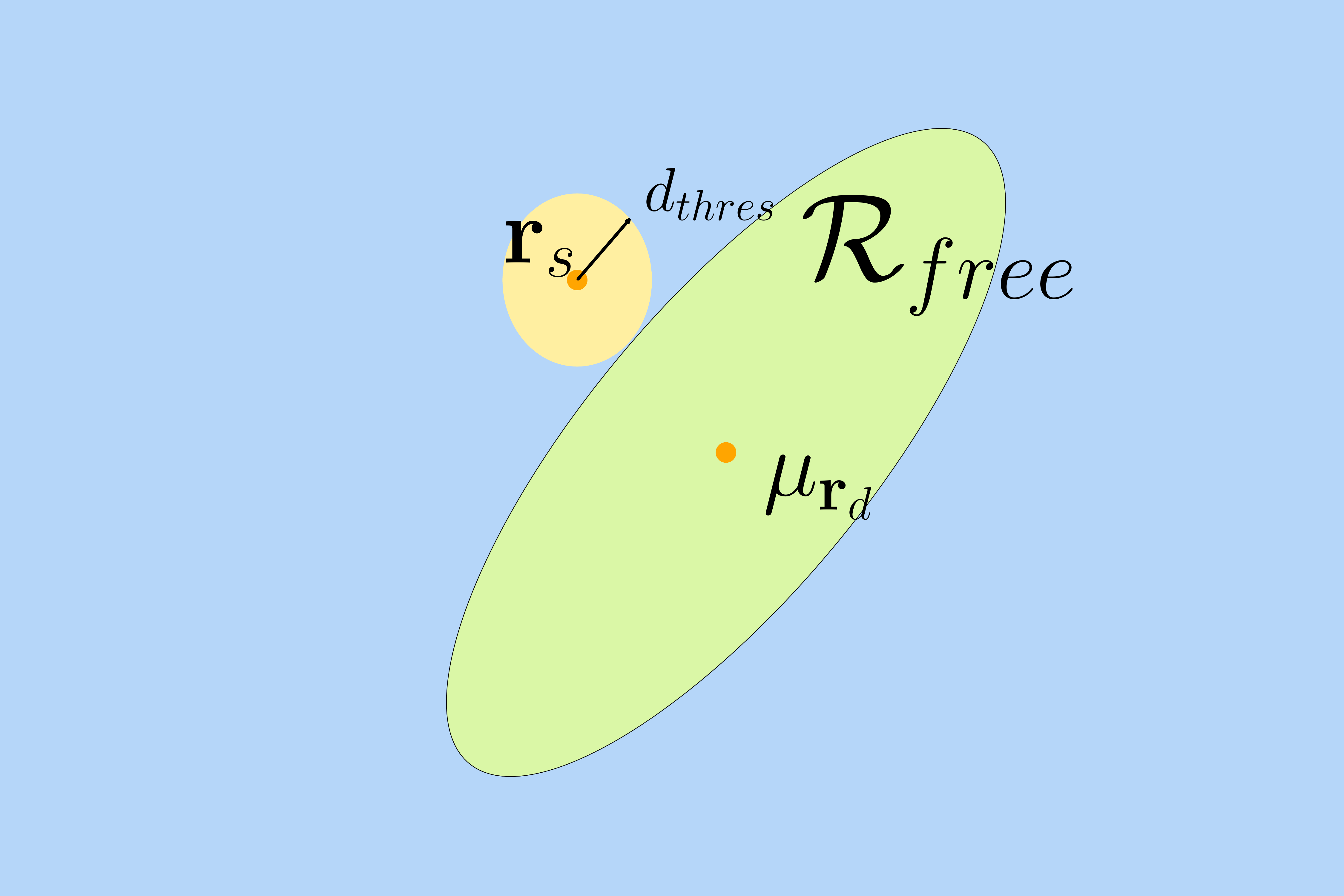}
    \caption{Under approximation of the collision-free set $\mathcal{R}_{free}^k$ as an ellipsoid. We assume the ellipsoidal set $\mathcal{R}_{free}^k$ is centered on mean $\mu_d^k$ of debris position $\mathbf{r}_d^k$.}
    \label{fig:safe_set}
\end{figure}
\noindent As we discussed before, we will use a convex safety cost function $l^k(\mathbf{r}_d^k)$. While the whole region outside of the collision threshold is a collision-free area (blue area in Figure~\ref{fig:original safe set}), this leads to a non-convex feasible set and safety cost function. Therefore, we under-approximate $\mathcal{R}_{free}$ as an ellipsoid centered on the debris mean position $\mu_d^k$ to convexify the safety cost function. Our ellipsoidal collision-free set and safety cost function for debris at time $k$ are as follows:
\begin{equation}\label{eq: ellipsoid safe set}
    \mathcal{R}_{free}^k = \big\{\mathbf{r} \in \mathbb{R}^3\ |\ l^k(\mathbf{r}) \leq 0 \big\}, \quad \text{where } l^k(\mathbf{r}) = (\mathbf{r}-\mu_d^k)^T E^k (\mathbf{r} - \mu_d^k) - 1, \quad E^k \succ 0, E^k \in \mathbb{R}^{3\times3}.
\end{equation}
where $E^k$ defines the shape of a collision-free set ellipsoid.

Then, our distributionally robust CVaR constraint can be reformulated as a function of the shape of the ellipsoidal safe set $E^k$ and covariance $\Sigma_d^k$~\cite{van2015distributionally}.
\begin{theorem} \label{thm:CVaR}
    For the random position vector $\mathbf{r}_d^k$, if collision-free set is defined as an ellipsoid $\mathcal{R}_{free}^k = \big\{\mathbf{r}\ |\ l^k(\mathbf{r}) = (\mathbf{r} - \mu_d^k)^T E^k (\mathbf{r} - \mu_d^k) -1 \leq 0 \big\}$, and $\mathcal{P}^k$ consists of all distributions of $\mathbf{r}_d^k$ that have mean $\mu_d^k$ and covariance $\Sigma_d^k$~\eqref{eqn: moment based ambiguity set}, then 
    \begin{equation} \label{eqn: CVaR reformulation}
        \sup_{\mathbb{P} \, \in \, \mathcal{P}^k} \textnormal{CVaR}_\varepsilon^{\mathbb{P}}  (l(\mathbf{r}_d^k)) = -1 + \frac{1}{\varepsilon}Tr\{\Sigma_d^k E^k\},
    \end{equation}
    where $Tr\{\cdot\}$ is the trace operator for matrices.
\end{theorem}
\begin{proof}
    See Proof of Corollary 1.3 in~\cite{van2015distributionally}.
\end{proof}
This closed-form solution~\eqref{eqn: CVaR reformulation} comes from solving dual problem of $\sup_{\mathbb{P} \, \in \, \mathcal{P}^k} \textnormal{CVaR}_\varepsilon^{\mathbb{P}}  (l(\mathbf{r}_d^k))$ with ambiguity set constraint~\eqref{eqn: moment based ambiguity set}. We can observe that the value of closed-form solution~\eqref{eqn: CVaR reformulation} increases as satisfying chance constraint becomes restrictive, such as small allowable collision probability $\varepsilon$, large debris position covariance $\Sigma_d^k$, or large $E^k$ which implies small collision-free set $\mathcal{R}_{free}$.

Finally, we can reformulate~\eqref{eqn: problem chance constraint} by using Theorem~\ref{thm:CVaR} for a tractable, conservative approximation of our distributionally robust chance-constrained optimal control problem.
\begin{subequations} \label{eqn: reformulated problem}
    \begin{align}
        \min_{\mathbf{u}_s^{0:K-1}} \mathcal{J} &:= \sum_{k\, =\, 0}^{K-1}  ({\mathbf{u}_s^k})^T R (\mathbf{u}_s^k) \label{eqn: objective}\\
        s.t. \quad & \mathbf{x}_{s}^{k+1} = f_s(\mathbf{x}_{s}^{k}, \mathbf{u}_{s}^{k}), \label{eqn: reformulated satellite dynamics}\\
            & \mathbf{x}_{d}^{k+1} = f_d(\mathbf{x}_{d}^{k}, \mathbf{w}_{d}^{k}), \label{eqn: reformulated debris dynamics}\\
            & \mathbf{x}_{d}^{0} \sim \mathcal{N}(\Bar{\mathbf{x}}_d^0, P_d^0) \\
           & -1 + \frac{1}{\varepsilon}Tr\{\Sigma_d^k E^k\} \leq 0 \quad \forall ~1 \leq k \leq K. \label{eqn: closed form CVaR constraint}
    \end{align}
\end{subequations}

\subsection{Constrained Cross-Entropy Method}

While we obtained closed-form approximation of chance constraint~\eqref{eqn: closed form CVaR constraint}, solving a constraint optimization problem~\eqref{eqn: reformulated problem} remains challenging due to the complex nonlinear dynamics of the satellites~\eqref{eqn: reformulated satellite dynamics} and debris~\eqref{eqn: reformulated debris dynamics}. We address the reformulated MPC problem~\eqref{eqn: reformulated problem} using a sampling-based optimization based on the Cross-Entropy Method (CEM)~\cite{de2005tutorial} to avoid the computationally expensive linearization or gradient computation of the nonlinear dynamics. CEM solves the optimization problem by iteratively sampling a set of candidate solutions from a predefined sampling distribution, evaluating their performance on objective, and updating the sampling distribution using a subset of the highest performing candidates. 

\begin{algorithm} [ht]
\caption{CEM optimization} \label{alg:main}
\hspace*{\algorithmicindent} \textbf{Input} Current satellite state $\mathbf{x}_s^0$, Initial debris state distribution $\mathbf{x}_{d}^{0} \sim \mathcal{N}(\Bar{\mathbf{x}}_d^0, P_d^0)$ \\
\hspace*{\algorithmicindent} \textbf{Output} Control input for satellite $\mathbf{u}_s^0$ 
\begin{algorithmic}[1]
\State  $\mu_d^{1:K}, \Sigma_d^{1:K} \leftarrow Uncertainty\_Propagation(\Bar{\mathbf{x}}_d^0, P_d^0)$ \Comment{Estimate mean and covariance of debris future position}
\For{iter = 1:max\_iteration}
    \State $\{ \mathbf{u}_s^{0:K-1} \}^{sampled} \leftarrow Sample(\mu_{CEM}, \sigma_{CEM})$ \Comment{Sample a set of control sequences candidates}
    \State $\{ \mathbf{x}_s^{0:K} \}^{sampled} \leftarrow dynamics(\mathbf{x}_d^0, \{ \mathbf{u}_s^{0:K-1} \}^{sampled})$ \Comment{Propagate sampled control sequences}
    \State Evaluate constraint~\eqref{eqn: closed form CVaR constraint} 
    \State Evaluate objective function $\mathcal{J}$~\eqref{eqn: objective}
    \If{All trajectories not feasible} \Comment{If all trajectory is not feasible,}
        \State Sort with $TrajectoryRisk$~\eqref{eqn: trajectory risk} \Comment{select control sequences with minimum risk}
        \State $\{ \mathbf{u}_s^{0:K-1} \}^{elite} \leftarrow SelectElite(\{ \mathbf{u}_s^{0:K-1} \}^{sampled})$ 
    \Else 
        \State $\{ \mathbf{u}_s^{0:K-1} \}^{feasible} \leftarrow SelectFeasible(\{ \mathbf{u}_s^{0:K-1} \}^{sampled})$ \Comment{Reject unsafe control sequences}
        \State Sort with $\mathcal{J}$~\eqref{eqn: objective} 
        \State $\{ \mathbf{u}_s^{0:K-1} \}^{elite} \leftarrow SelectElite(\{ \mathbf{u}_s^{0:K-1} \}^{feasible})$ \Comment{Select control sequences with minimum objectives}
    \EndIf
    \State $\mu_{CEM}, \sigma_{CEM} \leftarrow Update\big(\{ \mathbf{u}_s^{0:K-1} \}^{elite}\big)$
\EndFor
\State Get first control sequence in $\{ \mathbf{u}_s^{0:K-1} \}^{elite}$
\State Output first control input $u^*$
\end{algorithmic}
\end{algorithm}

Following~\cite{liu2020constrained, ryu2024integrating}, we use a variant of CEM to solve constrained MPC problems. First, we propagate debris trajectory and its initial uncertainty distribution based on its dynamics to obtain $\mu_d^{0:K}$ and $\Sigma_d^{0:K}$. Concurrently, we sample a set of candidate control sequences $\{ \mathbf{u}_s^{0:K-1} \}^{sampled}$ and propagate them through satellite dynamics~\eqref{eqn: satellite dynamics} to compute associated trajectories $\{ \mathbf{x}_s^{0:K} \}^{sampled}$. For each sampled trajectory, we evaluate the distributionally robust CVaR constraint~\eqref{eqn: closed form CVaR constraint} and reject trajectories that violate the constraints. After this rejection step, we define an elite set, $\{ \mathbf{u}_s^{0:K} \}^{elite}$, comprising control sequences that achieve optimal fuel consumption while satisfying the distributionally robust CVaR constraints. Finally, we update the parameters of the control sequence sampling distribution using this elite set and iterate the process until the maximum number of iterations is reached.

In the initial iterations of CEM, it may be challenging to identify feasible control sequences from the predefined sampling distribution. In such cases, it is crucial to update the sampling distribution to regions that produce safer control sequences. Therefore, we introduce \textit{Trajectory Risk} metric, defined as the discounted sum of~\eqref{eqn: closed form CVaR constraint}, which evaluates the overall risk associated with a control sequence:
\begin{equation}\label{eqn: trajectory risk}
    Trajectory\_Risk(\mathbf{x}_s^{0:K}, \mu_d^{1:K}, \Sigma_d^{1:K}) = \sum_{k=1}^K \gamma^k \left(-1 + \frac{1}{\varepsilon}Tr\{\Sigma_d^k E^k\}\right)
\end{equation}
where $\gamma$ is the discount factor. When there is no feasible control sequence that satisfies the distributionally robust CVaR constraint~\eqref{eqn: closed form CVaR constraint}, we select elite set $\{ \mathbf{u}_s^{0:K} \}^{elite}$ based on the minimum Trajectory Risk, which guides the control sequence sampling distribution towards safer control values. A summary of our MPC framework using CEM optimization is provided in Algorithm~\ref{alg:main}.

\section{Experiment Settings}\label{sec: experiment settings}

In this section, we will provide detailed settings of our simulation experiments. We will first provide context for the specific collision avoidance scenario we selected and the identification of the respective spacecraft. Then, as we test our controller with different uncertainty propagation methods, we will give brief overviews of the respective approaches.

\subsection{Satellite-Debris Collision Scenario}\label{subsection: collision scenario}

\subsubsection{Identification of Spacecraft and System Constraints}
We first identify a close approach scenario of satellite and debris through the SOCRATES Plus tool~\cite{kelso2006socrates} on CelesTrak. This tool utilizes STK’s Conjunction Analysis Tools~\cite{STK} and analyzes all objects in space, searching for all conjunctions within a distance of 5km at the Time of Closest Approach (TCA). It utilizes the Two-Line Element for all known orbiting spacecraft around the Earth and seeks out this conjunction and their TCA~\cite{kelso2006socrates}.

For our collision scenario, we selected one of the top 10 conjunctions by minimum range. We then retrieved the NORAD Catalog Number for the predicted conjunction and retrieved the Two-Line Element from the CelesTrak website\footnote{\href{https://celestrak.org}{https://celestrak.org}}. Then, we propagated the orbit of these two space objects in MATLAB using our satellite dynamics~\eqref{eqn: satellite dynamics} and debris dynamics~\eqref{eqn: debris dynamics} to verify the conjunction. 


\subsubsection{Collision Scenario Setup}
For this specific collision scenario, we utilized a predicted conjunction between a SpaceX Starlink spacecraft \emph{(NORAD: 52579)} and the THEA spacecraft \emph{(NORAD: 43796)}. Given that the THEA spacecraft is a 3U CubeSat ~\cite{THEAFCC} with no propulsion system, we can treat this as the debris in our experiment, and the SpaceX Starlink spacecraft as the satellite. These two spacecraft were predicted to have a conjunction range of  41 meters, according to the MATLAB simulation. Therefore, we define that there is a collision when satellite-debris distance is under $d_{threshold} = 100m$, and our controller has to take over to maintain this distance with probability $1-\varepsilon$.

In our simulation, we assume the mass of the satellite and debris are $300kg$ and $50kg$ respectively. The system noise of the debris dynamics, $\mathbf{w}_d$ in~\eqref{eqn: debris dynamics}, is modeled as white Gaussian noise, $\mathcal{N}(0, Q)$ with $Q = 0.05I$. The control input for our controller is constrained within the range $\mathbf{u}_{min} \leq \mathbf{u}_s \leq \mathbf{u}_{max}$ where $\mathbf{u}_{max} = [0.05, 0.05, 0.05] km/s^2$ and $\mathbf{u}_{min} = [-0.05, -0.05, -0.05] km/s^2$. While our discrete-time dynamics~\eqref{eqn: discrete dynamics} has a time step of $0.01s$, we compute the MPC controller every $1s$ and apply the same control for $100$ time steps for computation efficiency. Finally, we use a time horizon of $10s$ and a collision probability of $0.05$ for our controller. We note that our setup employs a relatively short MPC time horizon and higher $\mathbf{u}_{max}$ compared to conventional collision avoidance maneuvers. While this leads to higher $\Delta v$ as the system requires greater thrust for sudden trajectory changes to avoid collisions in the near future, it can highlight the variations in $\Delta v$ and minimum distance with respect to parameter changes, as discussed in Section~\ref{sec: results}.

\subsection{Uncertainty Propagation}
Since our distributionally robust constraint requires only the mean and covariance information of the debris position, our method has flexibility in the choice of uncertainty propagation method to use. In this experiment, we deploy our distributionally robust chance-constrained MPC with three different uncertainty propagation methods: (1) Linear Gaussian uncertainty propagation, (2) Unscented Transform, and (3) Monte Carlo simulations.

\subsubsection{Linear Propagation of Gaussian Uncertainty}
When initial uncertainty is Gaussian and the dynamics are linear, propagated uncertainty is also Gaussian~\cite{dutta2022convex}. Thanks to its computational simplicity and Gaussian assumption, it has been proposed as an effective uncertainty propagation method for short-term estimation purposes~\cite{chan2008spacecraft}. 
In our experiments, to use this method of uncertainty propagation, while we use nominal dynamics~\eqref{eqn: debris dynamics} to propagate the mean position of debris $\mu_d$, we need linearized dynamics to propagate the position covariance matrix $\Sigma_d$. We assume to have linearized discrete-time dynamics of debris as follows (See \cite{dutta2022convex} for details on linearization):
\begin{equation*}
    \delta\mathbf{x}_d^{k+1} = A \delta\mathbf{x}_d^{k} + \mathbf{w}_d, \quad  \mathbf{w}_d \sim \mathcal{N}(0, Q_{discrete})
\end{equation*}
where $A$ denotes the system matrix in linearized dynamics and $Q_{discrete}$ is the covariance of white system noise $\mathbf{w}_d$ in this discrete-time system, which can be computed from continuous-time dynamics in~\eqref{eqn: debris dynamics}~\cite{jacod2012discretization}.
Then, the propagation of debris state covariance $P_d$ can be achieved as
\begin{equation*}
    P_d^{k+1} = A P_d^k A^T + Q_{discrete}.
\end{equation*}
We can extract position error covariance $\Sigma$ from state error covariance $P$ since
\begin{equation*}
    P = 
    \begin{bmatrix}
        \Sigma & P_{\mathbf{r}\mathbf{v}} \\
        P_{\mathbf{r}\mathbf{v}}^T & P_{\mathbf{v}\mathbf{v}}
    \end{bmatrix}
\end{equation*}
where $P_{\mathbf{r}\mathbf{v}}$ represents the covariance between the position $\mathbf{r}$ and the velocity $\mathbf{v}$ and $P_{\mathbf{v}\mathbf{v}}$ denotes the covariance of velocity $\mathbf{v}$. Then, we can use these propagated mean and covariance of debris position for our distributionally robust chance constraint MPC.

\subsubsection{Unscented Transform}
While linear propagation provides a computationally efficient approach for analytical uncertainty propagation, it fails to capture higher-order terms in nonlinear dynamics. In contrast, the unscented transform is recognized for its superior ability to handle nonlinearity in system dynamics~\cite{vishwajeet2014nonlinear, ryu2021performance}, resulting in more accurate uncertainty estimates. The unscented transform leverages the propagation of sigma points, which are deterministically chosen sample points designed to precisely capture the mean and covariance of the distribution. These sigma points are propagated through the nonlinear dynamics, and the results are then used to estimate the mean and covariance of the propagated uncertainty.

Assuming that the initial state has mean $\Bar{\mathbf{x}}_d^0$ and covariance $P_d^0$, we can determine sigma points as
\begin{subequations}
    \begin{align*}
        &\mathbf{s}_0^0 = \Bar{\mathbf{x}}_d^0 \\
        &\mathbf{s}_i^0 = \Bar{\mathbf{x}}_d^0 + chol\left(\sqrt{nP_d^0} \right)_i, \quad \textnormal{for } i=1,\dots,n \\
        &\mathbf{s}_i^0 = \Bar{\mathbf{x}}_d^0 - chol\left(\sqrt{nP_d^0} \right)_{i-n}, \quad \textnormal{for } i=n+1,\dots,2n
    \end{align*}
\end{subequations}
where $n=6$ is the dimension of state vector $\mathbf{x} \in \mathbb{R}^6$ and $chol(\cdot)_i$ is the $i^\textnormal{th}$ column of $chol(\cdot)$, which is the Cholesky decomposition of its input. 

We propagate these sigma points through nonlinear debris dynamics~\eqref{eqn: debris dynamics} to compute propagated sigma points $\mathbf{s}^{1:K}_{0:2n}$. Then, we can estimate the mean and covariance of the debris state using these sigma points
\begin{subequations} \label{eqn: sample mean and cov}
    \begin{align}
        &\Bar{\mathbf{x}}_d^k = \frac{1}{2n+1} \sum_{i=0}^{2n} \mathbf{s}_i^k \\
        &P_d^k = \frac{1}{2n} \sum_{i=0}^{2n} (\mathbf{s}_i^k - \Bar{\mathbf{x}}_d^k) (\mathbf{s}_i^k - \Bar{\mathbf{x}}_d^k)^T,
    \end{align}
\end{subequations}
where we can extract debris position mean $\mu_d^k$ and covariance $\Sigma_d^k$ from debris state mean $\Bar{\mathbf{x}}_d^k$ and covariance $P_d^k$.

\begin{figure}[t] 
    \centering
    \begin{subfigure}[t]{0.45\textwidth}
        \centering
        \includegraphics[width=\textwidth]{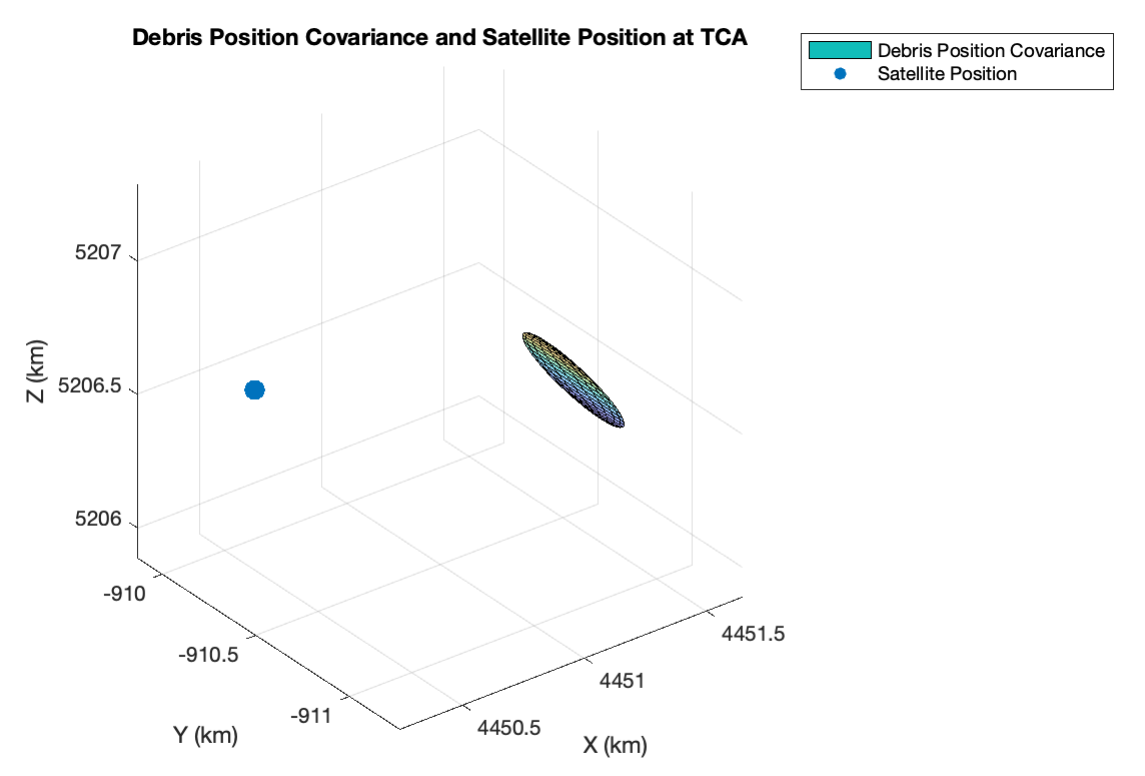} 
        \caption{Linear Gaussian propagation}
    \end{subfigure}
    \begin{subfigure}[t]{0.45\textwidth}
        \centering
        \includegraphics[width=\textwidth]{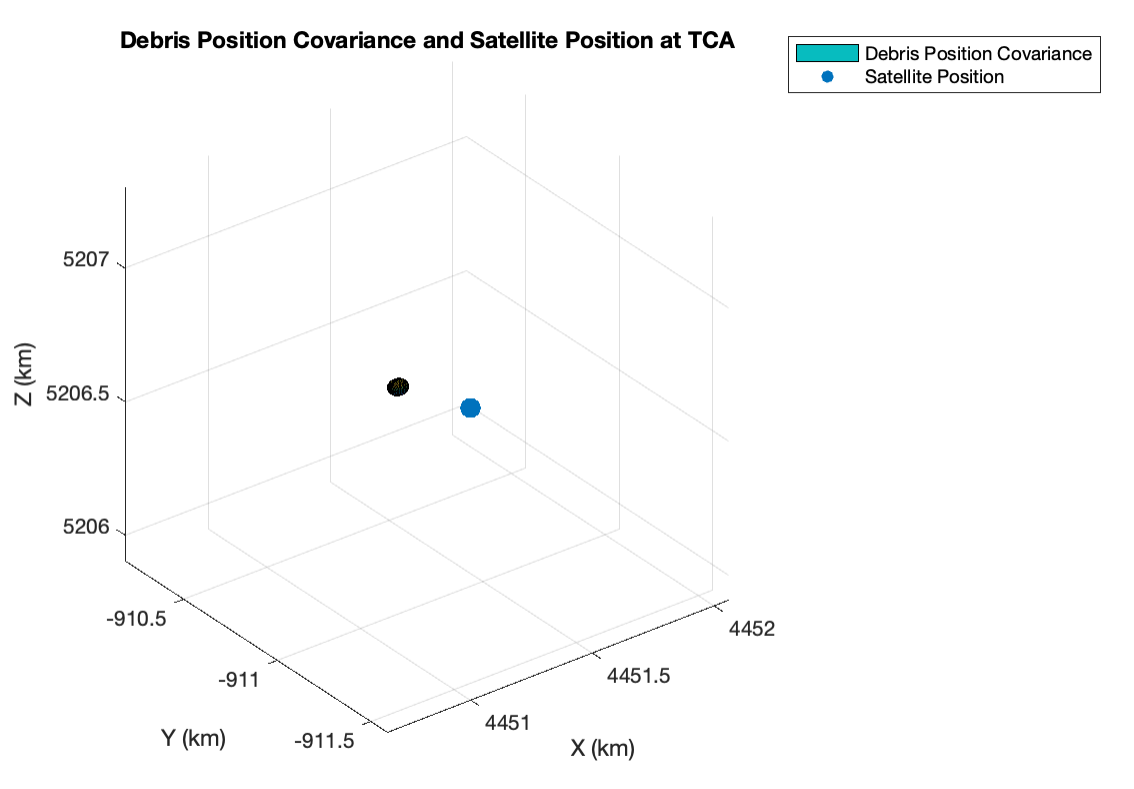} 
        \caption{Unscented transformation}
    \end{subfigure}

    \begin{subfigure}[b]{0.45\textwidth}
        \centering
        \includegraphics[width=\textwidth]{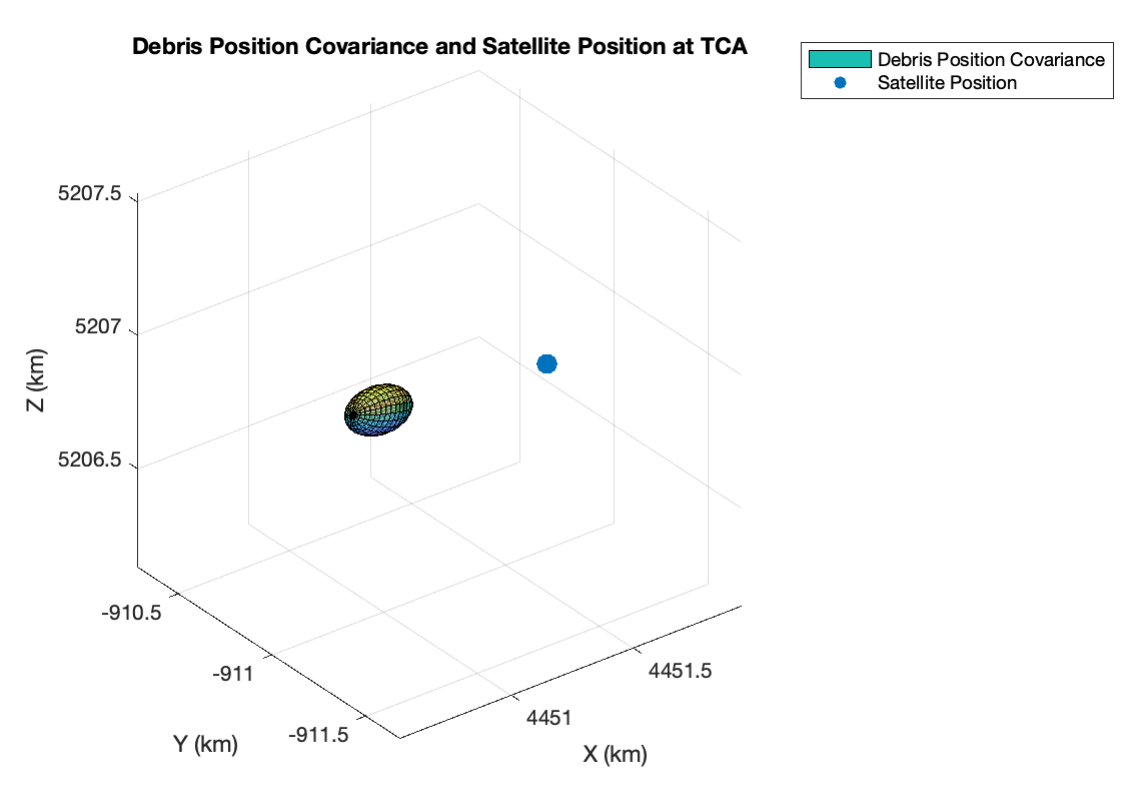} 
        \caption{Monte Carlo estimation}
    \end{subfigure}

    \caption{Debris position covariance and satellite position at Time of Closest Approach. We observe that while unscented transform under-estimate debris position uncertainty compared to Monte Carlo estimation, it can capture covariance shape similarly to Monte Carlo estimation. However, linear Gaussian propagation leads to a different shape of covariance which yields a different avoidance maneuver.}
    \label{fig: TCA uncertainty}
\end{figure}

\subsubsection{Monte Carlo Uncertainty Propagation}

While the Unscented Transform can capture up to second-order moments in nonlinear systems~\cite{wan2000unscented}, it fails to account for higher-order terms, which can result in divergence for cases far from the nominal orbit. In contrast, Monte Carlo simulation, which utilizes a significantly larger number of samples compared to the sigma points in the unscented transform, propagates these samples through the exact nonlinear dynamics, making it the most accurate method for orbit uncertainty propagation~\cite{de2010monte, sabol2011probability}. However, the primary limitation of Monte Carlo simulation lies in its computational cost, as it requires the propagation of a large number of trajectories~\cite{luo2017review}.

Typically, chance-constrained problems using Monte Carlo simulations rely on a sample approximation of the chance constraint~\cite{de2010monte, sabol2011probability}, where the ratio of samples encountering a collision to the total number of samples is evaluated. However, the drawback of this approach is the large sample size required for an accurate approximation of the chance constraint. Specifically, the required sample size for sample approximation of a chance constraint is proportional to $\log(1/\varepsilon)$~\cite{dagum2000optimal} when we want to enforce the chance constraint probability $1-\varepsilon$. This makes the method impractical for evaluating orbit collision probabilities, where the collision probability $\varepsilon$ is often very small. In contrast, we employ Monte Carlo simulation for estimating the mean and covariance of future uncertainty. This approach requires significantly fewer samples compared to directly approximating the chance constraint and is independent of the collision probability bound $\varepsilon$, making it suitable for evaluating low-probability events.

For the Monte Carlo simulation, we sample $N=50$ initial sample states $\mathbf{s}_{1:N}^0$ from initial state distribution $\mathcal{N}(\Bar{\mathbf{x}}_d^0, P_d^0)$. Then, we propagate sampled initial states using exact nonlinear debris dynamics~\eqref{eqn: discrete debris dynamics} to collect a set of future trajectories $\mathbf{s}^{1:K}_{1:N}$. Finally, we compute the mean and covariance of the debris state using sample mean and covariance estimation~\eqref{eqn: sample mean and cov}.

\section{Simulation Results} \label{sec: results}

In this section, we will verify our controller in simulation experiments. First, we give a qualitative analysis of our controller with different uncertainty propagation methods, showing how the quality of the uncertainty propagation method affects the result of our controller. Moreover, we provide the change of satellite-debris distance and required $\Delta v$ according to allowable collision probability $\varepsilon$ and the uncertainty in debris dynamics $\mathbf{w}_d$.

\subsection{Result of Uncertainty Propagation and Collision Avoidance Maneuver}

First, we give a qualitative analysis of different uncertainty propagation methods and their effects on collision avoidance maneuvers. We present debris covariance and position of the satellite at the Time of Closest Approach (TCA) in Figure~\ref{fig: TCA uncertainty}. We observe that while the unscented transform has under-estimated debris position uncertainty compared to Monte Carlo estimation, it still shows a similar covariance shape. This results in a similar satellite-debris configuration at TCA. Meanwhile, linear Gaussian propagation shows erroneous covariance estimation, which results in a different satellite-debris configuration from the other two methods. This result shows that linear Gaussian propagation has limitations when applied to highly nonlinear orbit dynamics, which can result in different maneuvers compared to the Monte Carlo estimate, which is known as the most accurate uncertainty propagation method.

\begin{figure}[t] 
    \centering
    \begin{subfigure}[t]{0.48\textwidth}
        \centering
        \includegraphics[width=\textwidth]{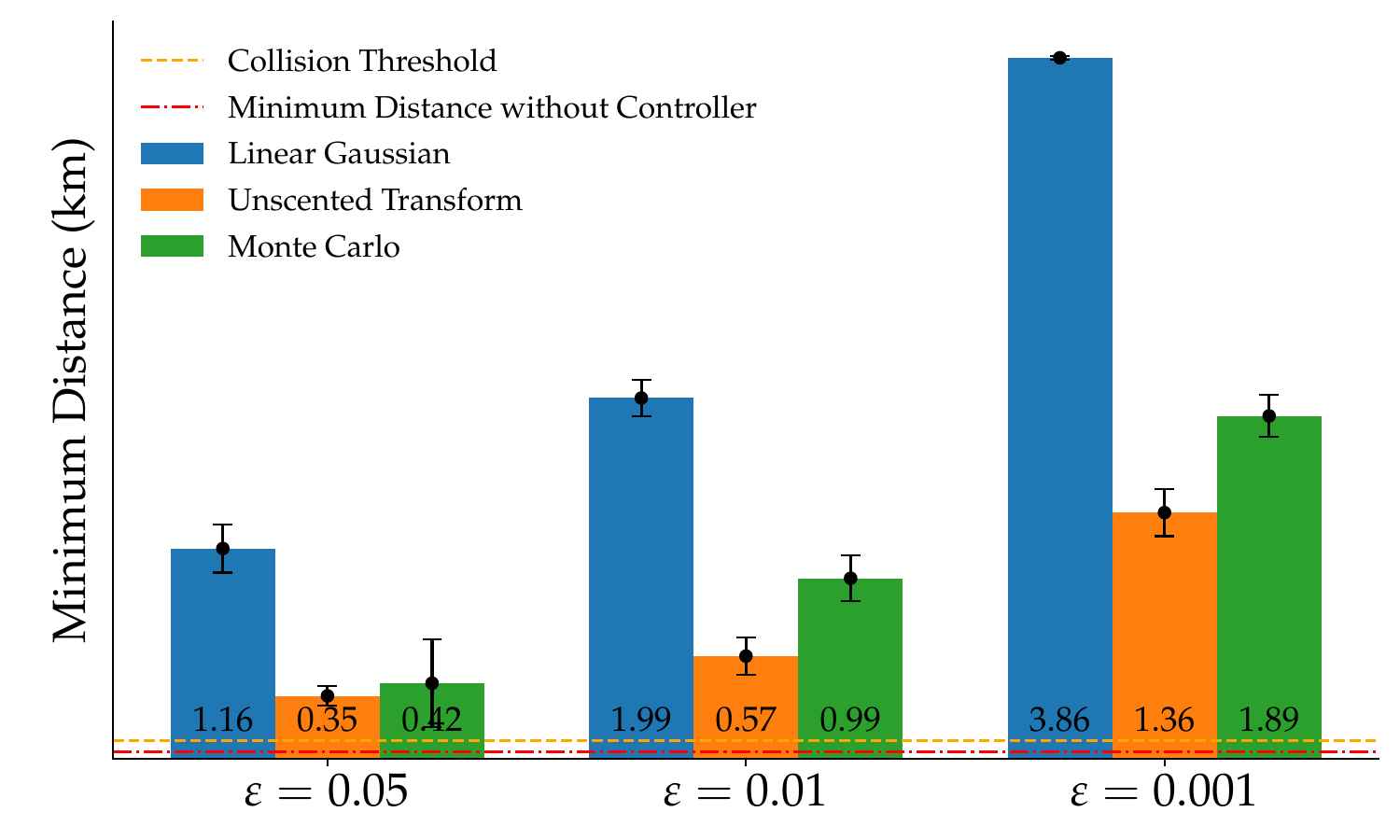} 
        \caption{Minimum distance between satellite and debris}
    \end{subfigure}
    \begin{subfigure}[t]{0.48\textwidth}
        \centering
        \includegraphics[width=\textwidth]{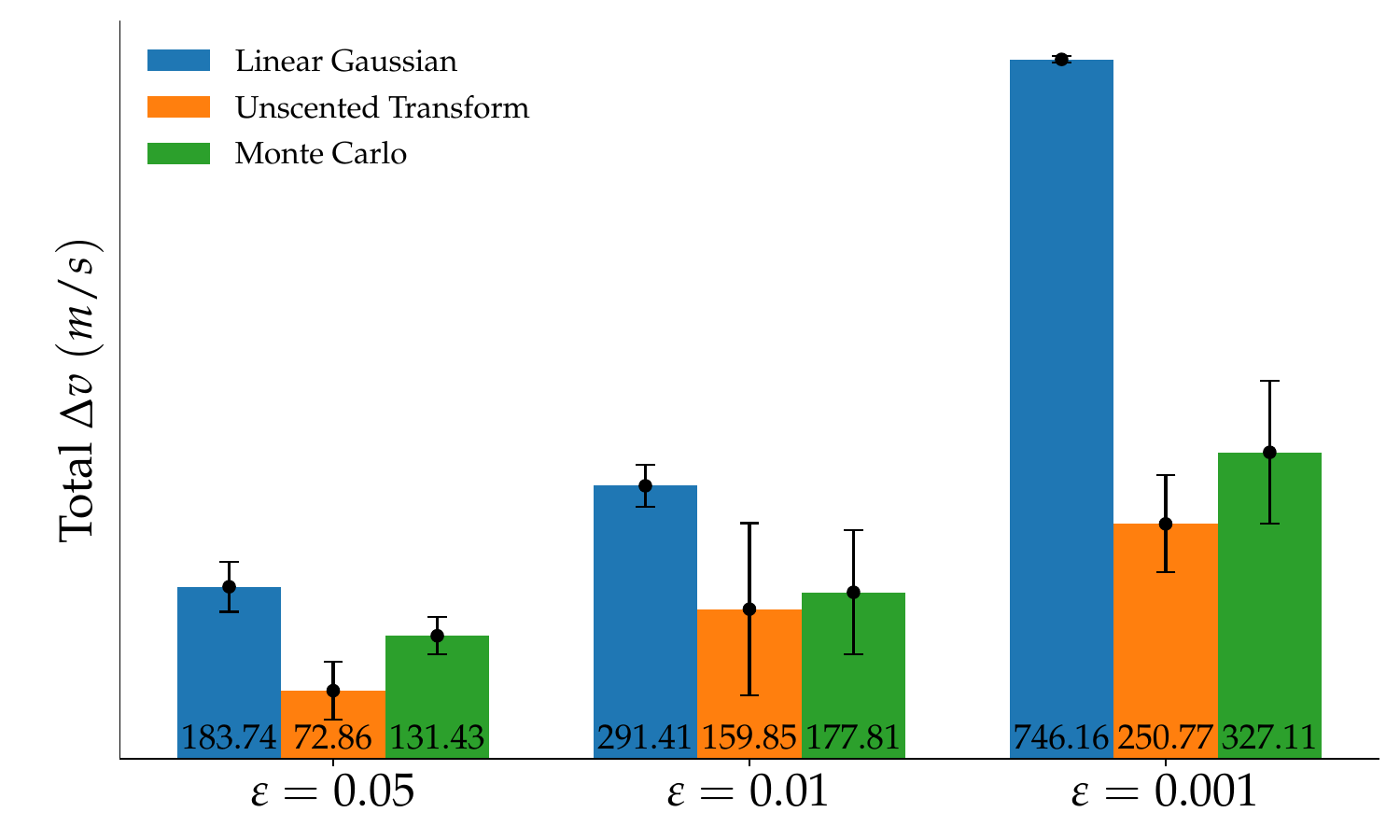} 
        \caption{Total $\Delta v$ in collision avoidance maneuver}
    \end{subfigure}

    \caption{Minimum distance between the debris and satellite during operation and total $\Delta v$ used for collision avoidance maneuvers with different uncertainty propagation methods and different chance constraint probability $\varepsilon$. We observe that enforcing lower collision probability leads to a conservative behavior, which maintains a larger minimum distance and incurs a larger $\Delta v$ cost.}
    \label{fig: different probability}
\end{figure}

\subsection{Effect of Chance Constraint Probability}
We report simulation results of our collision avoidance maneuver with different values for the upper bound on collision probability $\varepsilon$ in Figure~\ref{fig: different probability}. We run 10 experiments for each specific uncertainty propagation method and chance constraint probability $\varepsilon$ and report the mean and standard deviation of a minimum distance between the satellite and debris and the total $\Delta v$ used for maneuver.

In our simulation setup, we observe that as we reduce the allowable probability for collision, the controller tries to maintain further distance from the debris, which increases the cost by spending more fuel for avoidance maneuvers. This shows that our collision probability $\varepsilon$ can be an intuitive risk parameter when implementing our controller. Moreover, we observe that the controller applied with linear Gaussian propagation tends to use much more fuel to maintain overly conservative distances compared to Monte Carlo or unscented transform. This shows a limitation of linear Gaussian approximation, where inaccuracy of the uncertainty propagation can lead to inefficient maneuvers.

\begin{figure}[t] 
    \centering
    \begin{subfigure}[t]{0.48\textwidth}
        \centering
        \includegraphics[width=\textwidth]{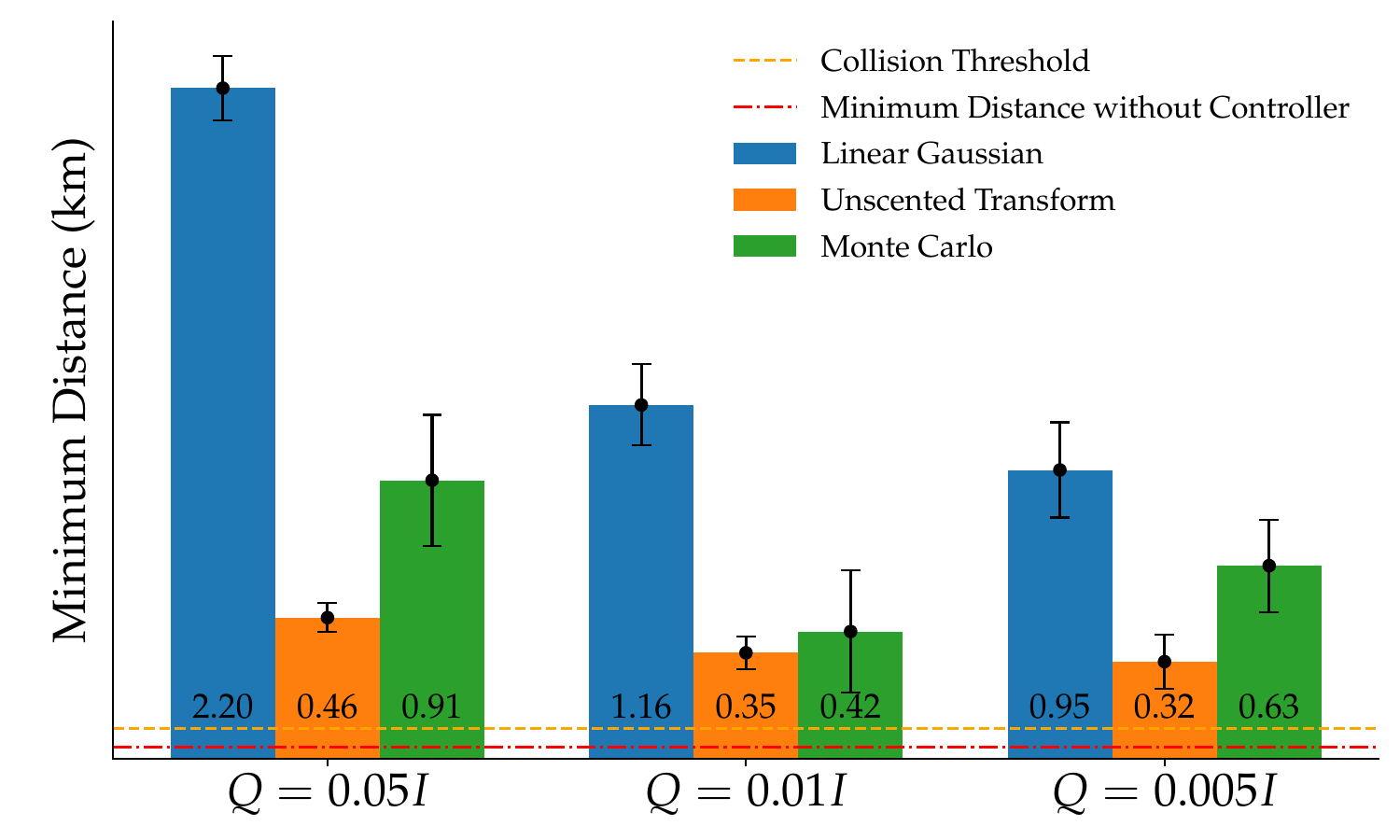} 
        \caption{Minimum distance between satellite and debris}
    \end{subfigure}
    \begin{subfigure}[t]{0.48\textwidth}
        \centering
        \includegraphics[width=\textwidth]{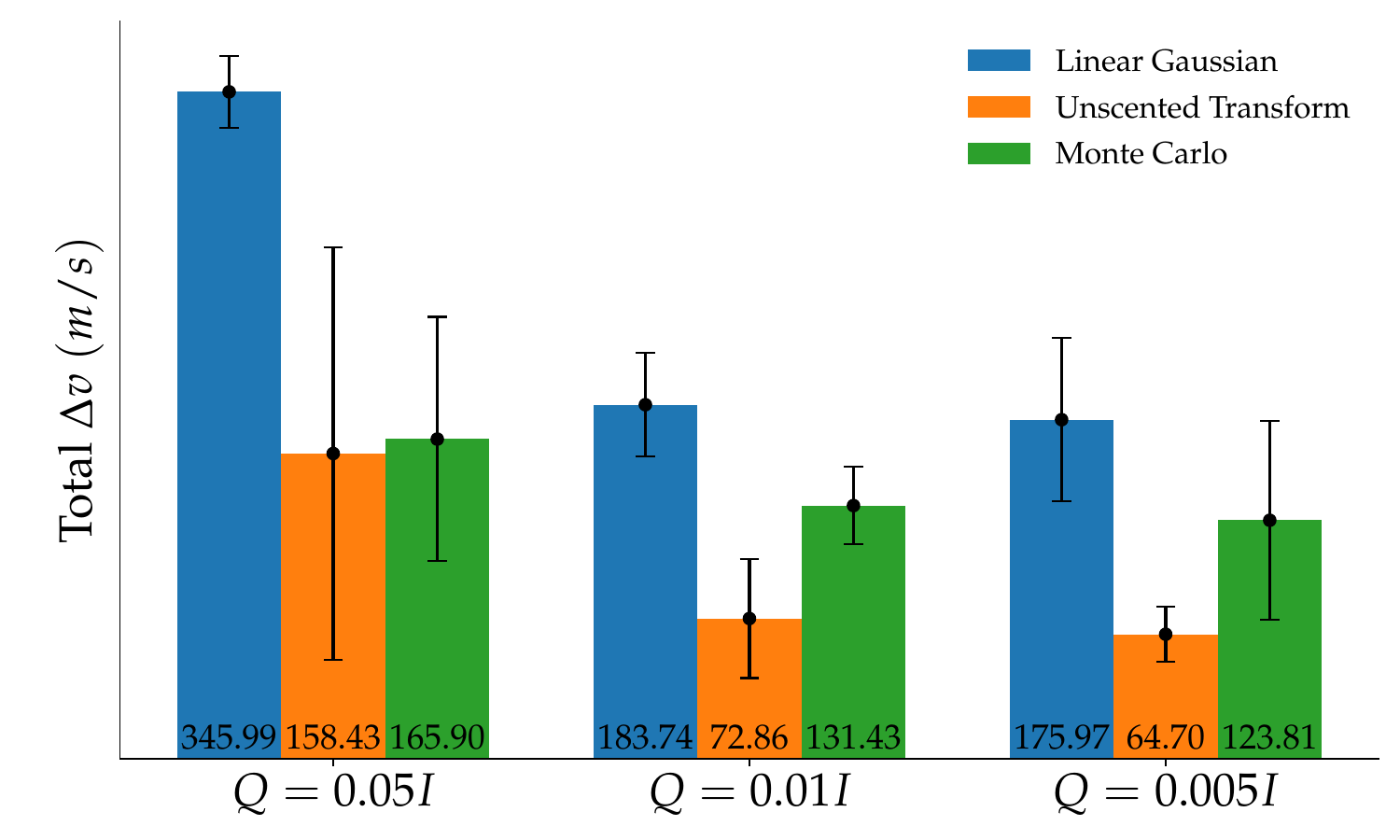} 
        \caption{Total $\Delta v$ in collision avoidance maneuver}
    \end{subfigure}
    \caption{Minimum distance between the debris and satellite during operation and total $\Delta v$ used for collision avoidance maneuvers with different debris uncertainty $Q$ for $\mathbf{w}_d \sim \mathcal{N}(0, Q)$. Our observations indicate that the controller tends to maintain a smaller minimum distance and consume less fuel as the uncertainty in debris dynamics decreases, corresponding to smaller values of $Q$.}
    \label{fig: different Q}
\end{figure}

\subsection{Effect of Uncertainty in Debris Dynamics}

We also evaluate the effect of debris uncertainty $\mathbf{w}_d$ in the avoidance maneuver in Figure~\ref{fig: different Q}, by conducting 10 experiments for each setup. We assume $\mathbf{w}_d$ is a white Gaussian noise $\mathcal{N}(0, Q)$ and compare the resulting minimum satellite-debris distance and total $\Delta v$ across different values of $Q$. 

Our results indicate that assuming less noisy debris dynamics reduces the required $\Delta v$, as the satellite can more accurately predict the debris' future position. Additionally, the satellite maintains a closer distance to the debris, reflecting the assumption that the debris position will deviate less from its nominal trajectory.

\section{Conclusion and Future Work}\label{sec: conclusion}

In this paper, we presented a collision avoidance algorithm that enforces chance constraints for spacecraft collision under debris trajectory uncertainty. To address the computational challenges of chance constraints under limited information about the uncertainty distribution, we approximated the chance constraint with a distributionally robust Conditional Value-at-Risk (CVaR) constraint. This approximation provides a sufficient condition for satisfying the chance constraint based on the mean and covariance estimates of the debris position. The collision avoidance problem was formulated within a Model Predictive Control (MPC) framework with a distributionally robust CVaR constraint and solved using the Cross-Entropy Method (CEM), a sampling-based optimization technique. Our approach successfully avoided collisions in a real-world-inspired scenario and demonstrated adaptability to varying parameters, including collision probability bounds and debris position uncertainty levels.

For future work, we aim to enhance the efficiency of collision avoidance maneuvers by replacing the sampling-based optimization method with more sophisticated solvers. Additionally, we plan to extend this work to address collision avoidance in scenarios involving multiple objects.

\section*{Acknowledgments}
This work is supported by the National Science Foundation, under grants
ECCS-2438314 CAREER Award, CNS-2423130, and CCF-2423131.
\bibliography{bibliograpy}

\end{document}